\documentclass[prl,aps,twocolumn,superscriptaddress]{revtex4}
\usepackage{latexsym}
\usepackage{graphicx}

\newcommand{\be}{\begin{equation}}
\newcommand{\ee}{\end{equation}}
\newcommand{\ba}{\begin{eqnarray}}
\newcommand{\ea}{\end{eqnarray}}
\newcommand{\baa}{\begin{eqnarray*}}
\newcommand{\eaa}{\end{eqnarray*}}

\def\be{\begin{equation}}
\def\ee{\end{equation}}
\def\bea{\begin{eqnarray}}
\def\eea{\end{eqnarray}}

\def\C60{A$_x$C$_{60}$}

\def\HgCu3{HgCa$_2$Cu$_3$O$_{8+y}$}
\def\HgCu4{HgBa$_2$Ca$_3$Cu$_4$O$_{10+y}$}
\def\TlCu{Tl$_2$Ba$_2$CuO$_{6+\delta}$}
\def\TlCu3{Tl$_2$Ba$_2$Ca$_2$Cu$_3$O$_{10+y}$}
\def\TlCu4{Tl$_2$Ba$_2$Ca$_3$Cu$_4$O$_{12+y}$}

\def\BiCu3{Bi$_2$Sr$_2$Ca$_{2}$Cu$_3$O$_y$}

\def\8LSCO{La$_{1.88}$Sr$_{.12}$CuO$_4$}
\def\110LNSCO{La$_{1.5}$Nd$_{0.4}$Sr$_{0.1}$CuO$_{4}$}
\def\stage4LCO{La$_{2}$CuO$_{4+\delta}$}
\def\Y248{YBa$_2$Cu$_4$O$_8$}

\def\NbSe2{NbSe$_2$}
\def\TaSe2{TaSe$_2$}
\def\TiSe2{TiSe$_2$}

\begin{document}

\title{Inter-layer Superconducting Pairing Induced $c$-axis Nodal Lines in Iron-based Superconductors}

\author{Yuehua Su}
\affiliation{Department of Physics, Yantai University, Yantai
264005, China}
\author{Chandan Setty}
\affiliation{Department of Physics, Purdue University, West
Lafayette, Indiana 47907, USA}
\author{Ziqiang Wang}
\affiliation{Department of Physics, Boston College, Chestnut Hill, Massachusetts 02467, USA}
\author{Jiangping Hu}
\email{hu4@purdue.edu}
\affiliation{Department of Physics, Purdue University, West
Lafayette, Indiana 47907, USA} \affiliation{Beijing National
Laboratory for Condensed Matter Physics, Institute of Physics,
Chinese Academy of Sciences, Beijing 100080,
China}

\begin{abstract}


A layered superconductor with a full pairing energy gap can be driven into a nodal superconducting (SC)  state by inter-layer pairing when the SC state becomes more quasi-3D. We propose that this mechanism is responsible for the observed nodal behavior in a class of iron-based SCs. We show that the intra- and inter-layer pairings generally compete and the gap nodes develop on one of the {\em hole} Fermi surface pockets as they become larger in the iron-pnictides. Our results provide a natural explanation of the c-axis gap modulations and gap nodes observed by angle resolved photoemission spectroscopy. Moreover, we predict that an anti-correlated $c$-axis gap modulations on the hole and electron pockets should be observable in the $S^{\pm}$-wave pairing state.

\end{abstract}

\maketitle

For the iron-based superconductors\cite{La,johnston} with a
complicated band structure, the symmetry of the order parameter in the
superconducting (SC) state \cite{hirsch} remains elusive. The
$S^\pm$-wave pairing symmetry, predicted by both
strong \cite{seo2008,Fang2011,huj,Yu2011} and weak coupling
theories \cite{mazin, Kuroki, WangF} based on the magnetic origin,
is a promising candidate and has been supported by many
experimental results \cite{hanaguri, nodeless1, huj, ZhangY2010}.
However, it has also been seriously challenged by the existence of gapless excitations or nodal behavior observed in some iron-based superconductors \cite{linenode1,linenode2,linenode3,linenode4,hhwen,syli}, in particular,
$BaFe_2As_{2-x}P_x$ \cite{iso5,isovalent3,isovalent5,isovalent9,dlfeng2010}
where some of the $\text{As}$ atoms are replaced by the $\text{P}$ atoms. A possible explanation of the nodal behavior has been suggested by the weak coupling approaches, such as the functional
renormalization group (FRG) technique \cite{dxy4,dxy3} and
random phase approximations (RPA) \cite{dxy5}. These calculations suggest that gap nodes can develop on the electron pockets when the detailed nesting properties vary among the hole pockets located at the $\Gamma$ and $M$ points, and the electron pockets located at the $X$ point of the unfolded Brillouin zone.  When the size of the hole pocket at M point decreases, the SC gap on the electron pockets becomes increasingly anisotropic and eventually gap nodes emerge. The reduction of the M hole pocket can be achieved by either increasing electron doping or by tuning the pnictogen height through replacing $\text{As}$ by
$\text{P}$\cite{isovalent9,dxz}.

Recently, nodes in the gap function dispersion along in the $c$-axis ($c$-axis nodal lines) have been observed directly by ARPES in $BaFe_2As_{1.7}P_{0.3}$ \cite{dlfeng2011}. However, the weak coupling theories cannot explain the observed
nodal behavior for the following reasons. First, the
observed nodes are on the hole pockets, not on the electron pockets. Second, in contrast to
LDA calculations \cite{dxy5}, the $\text{P}$ substitution in these
materials does not push the hole-like band near M to sink below the Fermi
surface \cite{dlfeng2010,isovalent3}. Instead, as the substitution
increases, the M hole pocket and the X electron pockets barely
change while the hole pockets near the $\Gamma$ point at the zone center ($k_z=0$), which have large c-axis dispersions, accommodate the additional holes. As a result, with increasing P substitution, one of the two $\Gamma$ hole pockets
grows increasingly larger. The size of this hole pocket at the $Z$
point ($k_z=\pi$) can even be larger than the size of the largest hole
pocket in $KFe_2As_2$, the most hole-doped iron-based
superconductors known today.  These properties point to a non-rigid band picture
under the``iso-valent" doping and completely violate the assumption
of the band structure taken in the above weak coupling theories.

In this Letter, we suggest that the observed nodal behavior originates from
the inter-layer pairing and the reduction of the
intra-layer SC pairing gap due to the increase of the
size of the hole pockets. This proposal consistently explains the
$c$-axis modulation of the SC gaps observed in optimally hole-doped
$Ba_{1-x}K_xFe_2As_2$\cite{ZhangBK,hding} and the nodal behaviors
in  $BaFe_2As_{2-x}P_x$\cite{dlfeng2011}. It suggests that the
gap modulations along the $c$-axis are directly related to the $c$-axis
band dispersion. Our calculation also reveals that the inter-layer
SC pairing, in general, competes with the intra-layer SC pairing
in these quasi-two dimensional materials, a possible reason why
the highest $T_c$ is not achieved in the 122-family ($AFe_2As_2$) but
in the 1111-family ($AOFeAs$) of the iron pnictides\cite{ren,xchen} since
the former is more three dimensional \cite{schi,huiqiu,johnston}. Moreover, we predict that the $S^\pm$-wave
pairing symmetry should result in an anti-correlation of the SC gap values between  the hole and electron pockets along the $c$-axis as a function of $c$-axis
momentum. This property, if observed, can serve as a direct experimental evidence for the $S^\pm$-wave pairing symmetry.

{\it Model} We construct a three-orbital model which includes
the $d_{xz},d_{yz}$ and $d_{z^2}$ orbitals to study the physics.
Experimentally, the large  $c$-axis dispersion is only observed in
one of the hole pockets near the $\Gamma$ point which is mainly composed
of $d_{xz,yz}$ orbitals \cite{YZhang}. The increase of the $c$-axis
dispersion upon \text{P} doping is mainly due to the increase of
the mixture of the $d_{z^2}$ orbital into this hole
pocket \cite{dlfeng2010}. This has been shown by both polarized
ARPES experiments \cite{dlfeng2010} and numerical
calculations \cite{dxz,xdai}. The ARPES
experiments \cite{dlfeng2010,dlfeng2011} show that the hole pocket with
the large $c$-axis dispersion has even symmetry with respect to the
reflection of the $\Gamma-M$ mirror plane and, with increasing $P$
doping, the band mainly attributed to the $d_{z^2}$ orbital moves
closer and closer to the Fermi energy so that the weight of $d_{z^2}$
on the Fermi surface increases. This picture is consistent with the
symmetry analysis since the $d_{z^2}$  orbital is also symmetric
with respect to the reflection of the $\Gamma-M$ mirror plane. The
model we construct captures all the above essential experimental
results and can still achieve high analytical tractability. We will
also show that the results for the $c$-axis properties derived from this
model are rather generic.

\begin{figure}[thbp]
\includegraphics [angle=0,width=0.45\columnwidth,clip]{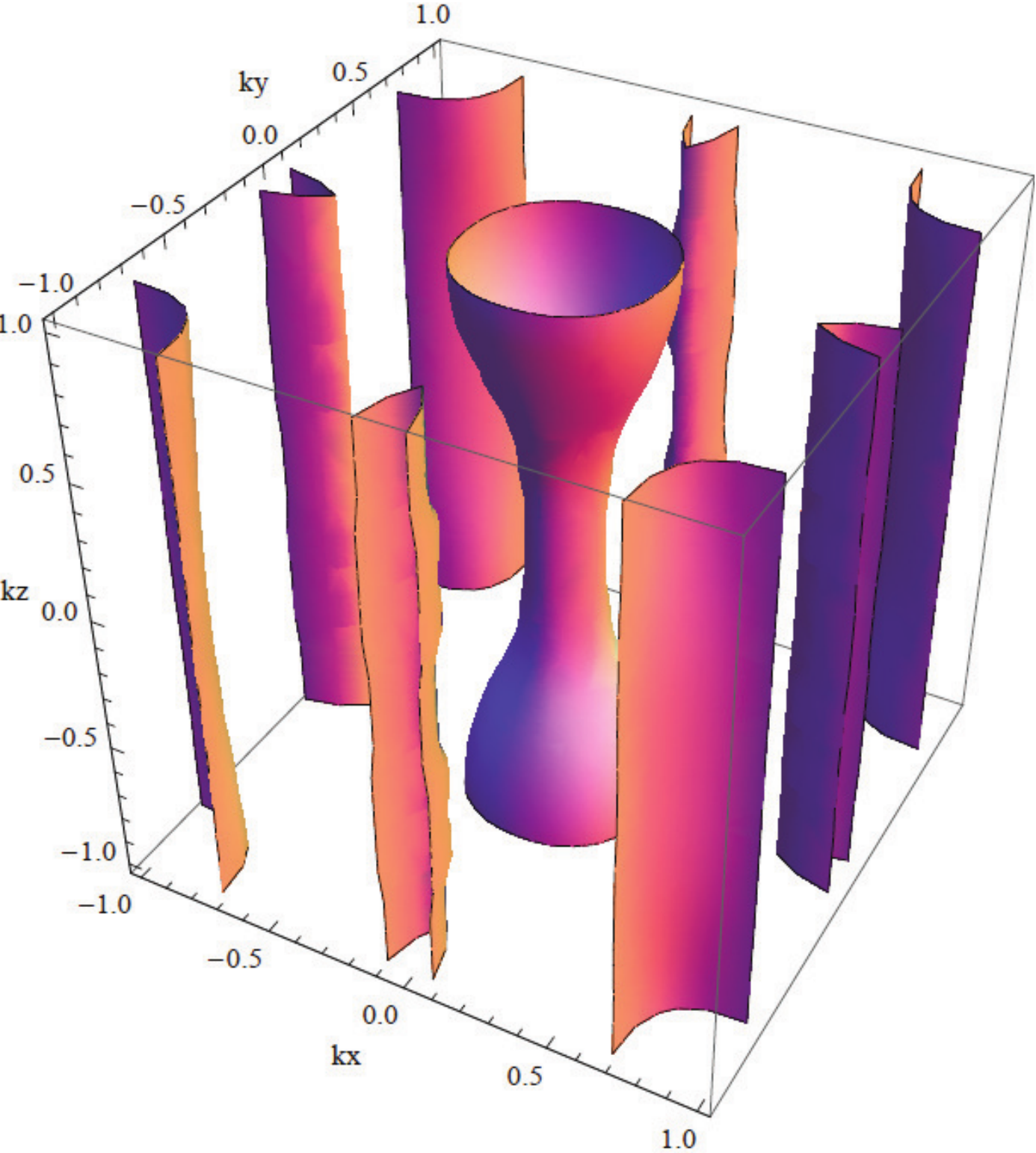}
\includegraphics [angle=0,width=0.45\columnwidth,clip]{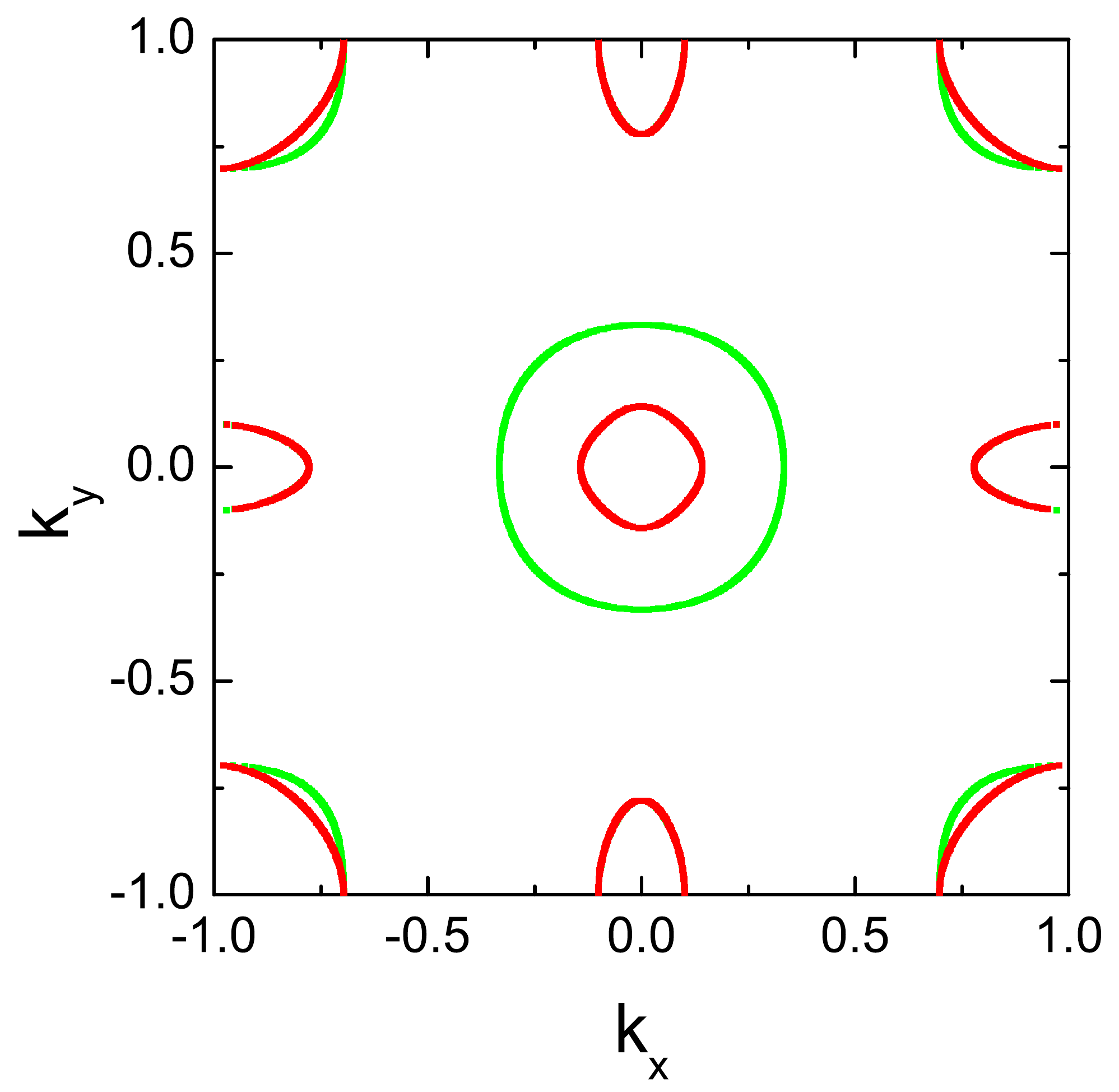}
\caption{ 3-D Fermi surfaces and Contour plot of Fermi surfaces in the extended
three-band model Eq.(\ref{eqn3.3}) ($k_z=0$ for the read line
and $k_z=\pi$ for the green line).
 } \label{fig3.1}
\end{figure}

The Hamiltonian of our model includes two parts
$
H=H_t + H_{I} ,
$
where $H_t$ is the kinetic energy and $H_I$ is the pairing interaction. $H_t$ is given by
\begin{equation}
H_t = \sum_{\mathbf{k},\alpha\beta\sigma}
\varepsilon_{\mathbf{k}\alpha\beta}
d_{\mathbf{k}\alpha\sigma}^{\dag}d_{\mathbf{k}\beta\sigma},
\label{eqn3.3} \nonumber
\end{equation}
where $\alpha,\beta =1,2,3$ label the 3d electrons in
the $d_{xz}$, $d_{yz}$ and $d_{z^2}$ orbital respectively.  The
form of the kinetic energy and the hopping parameters are constructed by
extending the two-orbital model \cite{raghu} to achieve the $c$-axis
dispersion that matches well the experimental results. Their
explicit forms are
\begin{eqnarray}
&& \varepsilon_{\mathbf{k},11} = -2 t_1 \cos k_x -2t_2 \cos k_y -
4
t_3 \cos k_x \cos k_y -\mu ,  \nonumber \\
&& \varepsilon_{\mathbf{k},22} = -2 t_2 \cos k_x -2t_1 \cos k_y -
4
t_3 \cos k_x \cos k_y -\mu ,  \nonumber \\
&&\varepsilon_{\mathbf{k},33} = 2 t_z \left(\cos k_x + \cos k_y\right) -\mu_3 ,  \nonumber \\
&&\varepsilon_{\mathbf{k},12} = - 4 t_4 \sin k_x \sin k_y ,  \nonumber \\
&&\varepsilon_{\mathbf{k},13} = -t_{xz} \left(1-\cos
k_z\right)\left(\cos k_x + \cos k_y\right) \sin k_x ,  \nonumber \\
&&\varepsilon_{\mathbf{k},23} = t_{xz} \left(1-\cos
k_z\right)\left(\cos k_x + \cos k_y\right) \sin k_y .  \nonumber
\end{eqnarray}
The c-axis dispersion is induced by the coupling between the $d_{xz,yz}$ and $d_{z^2}$ orbitals.
The later is taken to be below the Fermi level. In the explicit form of the coupling between these orbitals,
$\varepsilon_{\mathbf{k},13}$, we have taken into account both the lattice symmetry requirements and
the experimental observations. By taking the following  parameters
$t_1 =-1, t_2 = 1.55, t_3=t_4=-0.85, t_z=1, t_{xz} = 0.8, \mu
= 1.77, \mu_3 = 5$, we find that the model describes well the
3-dimensional Fermi surfaces measured experimentally \cite{dlfeng2010,dlfeng2011} as
shown in Fig. \ref{fig3.1}.

$H_I$, the SC pairing interaction, generally includes the following terms,
\begin{eqnarray}
H_{intra,1} & = &
-V_1\sum_{\mathbf{kk'},\alpha=1,2}\phi^{(1)}_{\mathbf{k}}\phi^{(1)}_{\mathbf{k'}}
d_{\mathbf{k}\alpha\uparrow}^{\dag}d_{\mathbf{-k}\alpha\downarrow}^{\dag}
d_{\mathbf{-k'}\alpha\downarrow}d_{\mathbf{k'}\alpha\uparrow} \nonumber \\
H_{inter,2}& =&
-V_2\sum_{\mathbf{kk'},\alpha=1,2}\phi^{(2)}_{\mathbf{k}}\phi^{(2)}_{\mathbf{k'}}
d_{\mathbf{k}\alpha\uparrow}^{\dag}d_{\mathbf{-k}\alpha\downarrow}^{\dag}
d_{\mathbf{-k'}\alpha\downarrow}d_{\mathbf{k'}\alpha\uparrow} \nonumber\\
H_{inter,3}&
=&-V_3\sum_{\mathbf{kk'}}\phi^{(3)}_{\mathbf{k}}\phi^{(3)}_{\mathbf{k'}}
d_{\mathbf{k},3\uparrow}^{\dag}d_{\mathbf{-k},3\downarrow}^{\dag}
d_{\mathbf{-k'},3\downarrow}d_{\mathbf{k'},3\uparrow} \nonumber \\
H_{inter,4}& =&
-V_4\sum_{\mathbf{kk'}}\phi^{(4)}_{\mathbf{k}}\phi^{(4)}_{\mathbf{k'}}
(d_{\mathbf{k},1\uparrow}^{\dag}d_{\mathbf{-k},1\downarrow}^{\dag}
d_{\mathbf{-k'},3\downarrow}d_{\mathbf{k'},3\uparrow}\nonumber \\
&
&+d_{\mathbf{k},2\uparrow}^{\dag}d_{\mathbf{-k},2\downarrow}^{\dag}
d_{\mathbf{-k'},3\downarrow}d_{\mathbf{k'},3\uparrow} + h.c.).
\nonumber
\end{eqnarray}
The first term $H_{intra,1}$ describes the intra-layer pairing
while the rest three terms describe the different inter-layer
pairing interactions. $H_{inter,2}$ accounts for the inter-layer pairing
interaction between the $d_{xz}$ and $d_{yz}$ orbitals, as does
$H_{inter,3}$ for the $d_{z^2}$ orbital. The last term describes the
inter-layer interaction between the $d_{z^2}$ and $d_{xz,yz}$
pairs.  We take $\phi^{(1)}_{\mathbf{k}}$ to be the intra-layer $S^{\pm}$-wave
paring function $\phi^{(1)}_{\mathbf{k}}=\cos k_x \cos k_y$ and
$\phi^{(2)}_{\mathbf{k}}=(\cos k_x + \cos k_y)\cos k_z$, and
$\phi^{(3,4)}_{\mathbf{k}}=\cos k_z$. These choices rely on the assumption that the SC
pairing in iron-based superconductors is rather short-ranged. The
form of $\phi^{(1)} $ has been proposed in the models based on
local magnetic exchange couplings \cite{seo2008,Fang2011,huj,Yu2011} and it has been shown that the
form factor is consistent with current experimental results \cite{nodeless1}. The
form of $\phi^{(2)}$ has been proposed in \cite{hding}, which can be
obtained from the existence of AFM exchange couplings between the
layers \cite{daipc,daipc2}. The form of $\phi^{(3,4)}$ and the corresponding $V_{3,4}$
pairing interactions can be understood  as the inter-layer pairing is between two adjacent  layers and the pairing symmetry
is s-wave. $V_3$ describes the inter-layer pairing
for the $d_{z^2}$ orbital. The $V_4$ term, which describes the  coupling between two inter-layer pairings of  two different orbitals, can be understood in the following way. Since the
c-axis hopping term in $H_t$ describes the hopping between two
adjacent layers and the $d_{z^2}$ is below the Fermi level, the
second order perturbation through such hopping would generically
produce $V_4\propto
\frac{t_{xz}^2}{\mu_3}$. 

In the self-consistent mean-field theory for the SC state,  $H_I$  becomes
\begin{eqnarray}
H_{BCS} = -\sum_{\mathbf{k}\alpha} \Delta_{\alpha}(\mathbf{k})
d_{\mathbf{k}\alpha\uparrow}^{\dag}d_{\mathbf{-k}\alpha\downarrow}^{\dag}
+ h.c. , \label{eqn3.6}
\end{eqnarray}
where $\Delta_{1}(\mathbf{k})=\Delta^{(1)}_{1}
\phi^{(1)}_{\mathbf{k}} + \Delta^{(2)}_{1} \phi^{(2)}_{\mathbf{k}}
+ \Delta^{(4)}_{3} \phi^{(4)}_{\mathbf{k}} $,
$\Delta_{2}(\mathbf{k})=\Delta^{(1)}_{2} \phi^{(1)}_{\mathbf{k}}
+\Delta^{(2)}_{2} \phi^{(2)}_{\mathbf{k}}+ \Delta^{(4)}_{3}
\phi^{(4)}_{\mathbf{k}} $ and
$\Delta_{3}(\mathbf{k})=\Delta^{(3)}_{3} \phi^{(3)}_{\mathbf{k}}
+(\Delta^{(4)}_{1} + \Delta^{(4)}_{2}) \phi^{(4)}_{\mathbf{k}} $.
Here $\Delta^{(n)}_{\alpha}$ is defined by
\begin{eqnarray} \Delta^{(n)}_{\alpha} = V_n
\sum_{\mathbf{k}}\phi^{(n)}_{\mathbf{k}} \langle
d_{\mathbf{-k}\alpha\downarrow}d_{\mathbf{k}\alpha\uparrow}
\rangle . \label{eqn3.7}
\end{eqnarray}

\begin{figure}[thbp]
\includegraphics [angle=0,width=0.65\columnwidth,clip]{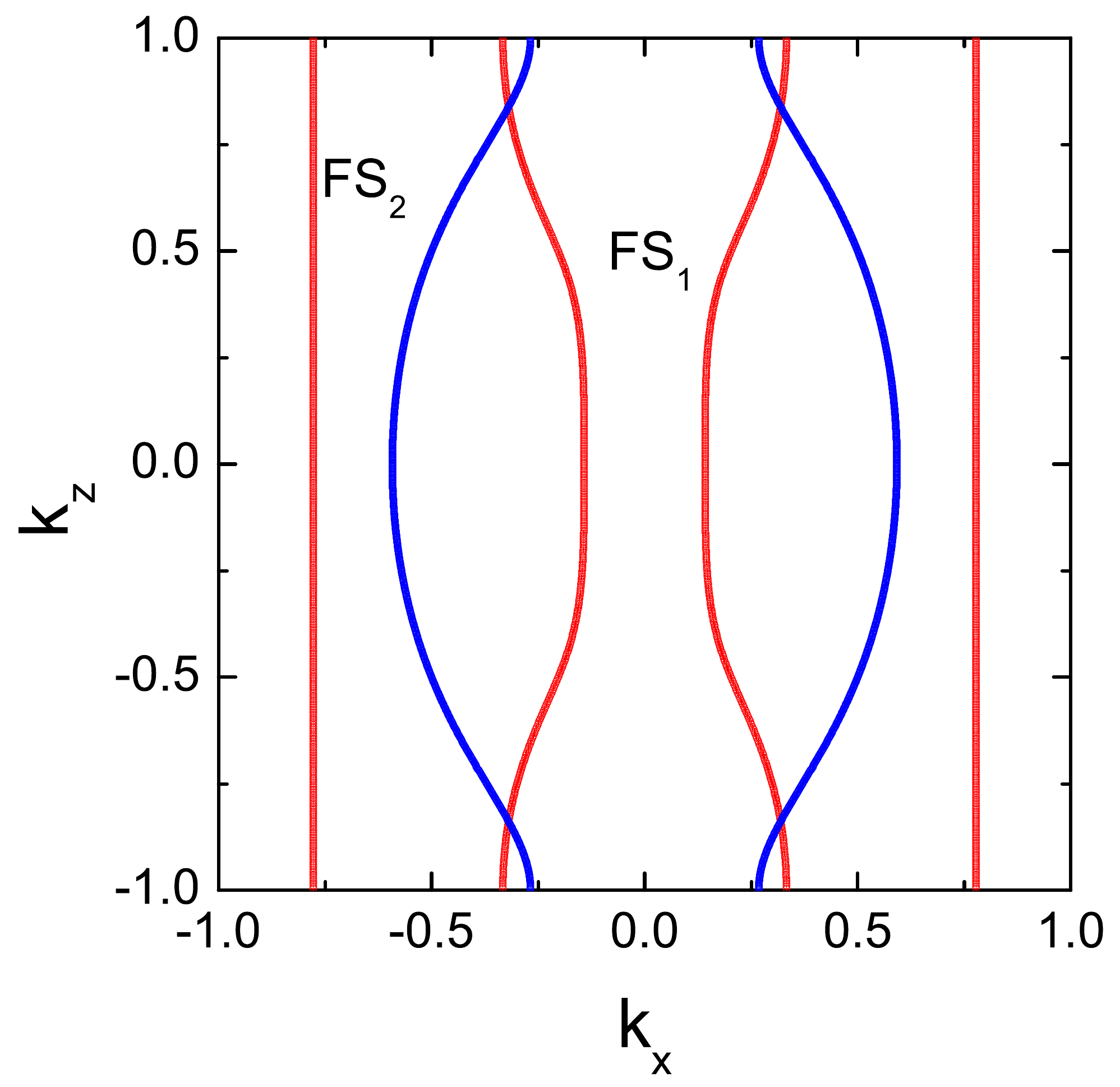}
\caption{ A cartoon plot for the gap nodal picture in the momentum
space $(k_x,0,k_z)$. The blue lines are the contour lines for zero
gap value of Eq. (\ref{eqn3.1}) with $\delta_z=0.4$ and
$\lambda=0$. FS$_1$ and FS$_2$ are defined in Fig. \ref{fig3.1}.
The intercept between the blue and the red lines produce the
nodes in the hole pocket. } \label{fig3.2}
\end{figure}

{\it Results} Before we present a full numerical solution for the
above Hamiltonian, we first discuss the simple physical picture for the
generation of the nodal points in the gap function on the hole pocket. In the
above model, if we consider the general intra-orbital pairing form of the $d_{xz,yz}$  orbitals,  the $k_z$-dependent SC gap can be written as
\begin{equation}
\Delta(\mathbf{k}) = \Delta_{0} \left[ \cos k_x \cos k_y +
\delta_z \left(\lambda+\cos k_x + \cos k_y \right) \cos k_z
\right] , \label{eqn3.1}
\end{equation}
where the first term represents the $S^\pm$ pairing and the second term represents the inter-layer pairing with the $S$-wave
pairing symmetry. In the first order approximation, this intra-orbital pairing  roughly determines the SC gap since it dominates as we will show later. This form indicates that the inter-layer
pairing is between the two neighboring layers. The gap zero points develop as $\delta_z$ increases. As
shown in Fig. \ref{fig3.2}, when $\delta_z$ reaches a certain value, the contour of the gap zeroes will cross the
Fermi surface at the points near $k_z=\pm \pi$, which leads to the nodal behavior.

\begin{figure}[tbp]
\includegraphics [angle=0,width=0.9\columnwidth,clip]{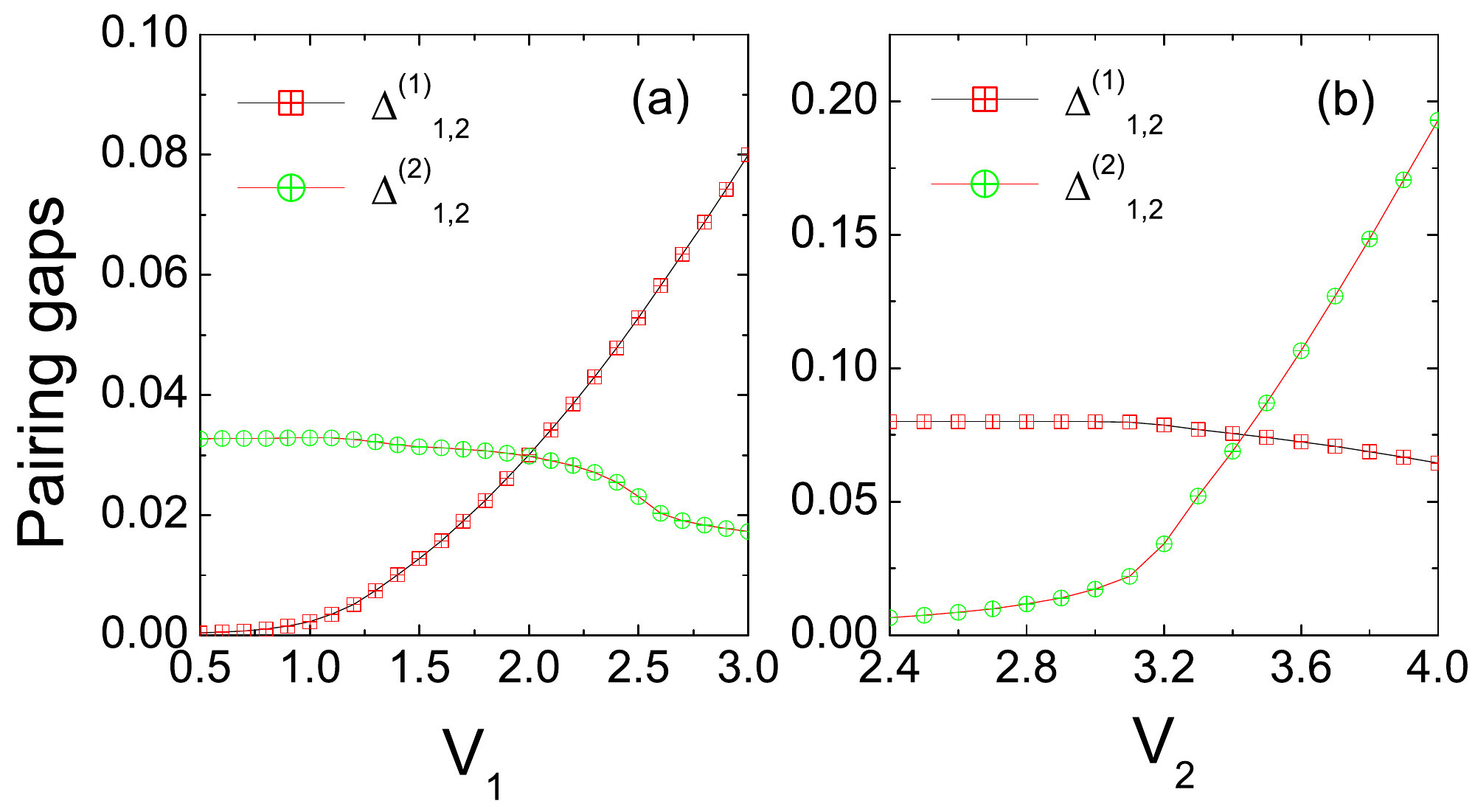}
\caption{ Gaps versus pairing interactions. Interaction parameters
in (a) are defined as $V_2=3, V_3=V_4=0$ and (b) $V_1=3,
V_3=V_4=0$. $\Delta^{(n)}_{\alpha}$ is defined in Eq.
(\ref{eqn3.7}). } \label{fig3.3}
\end{figure}

\begin{figure}[tbp]
\includegraphics [angle=0,width=0.8\columnwidth,clip]{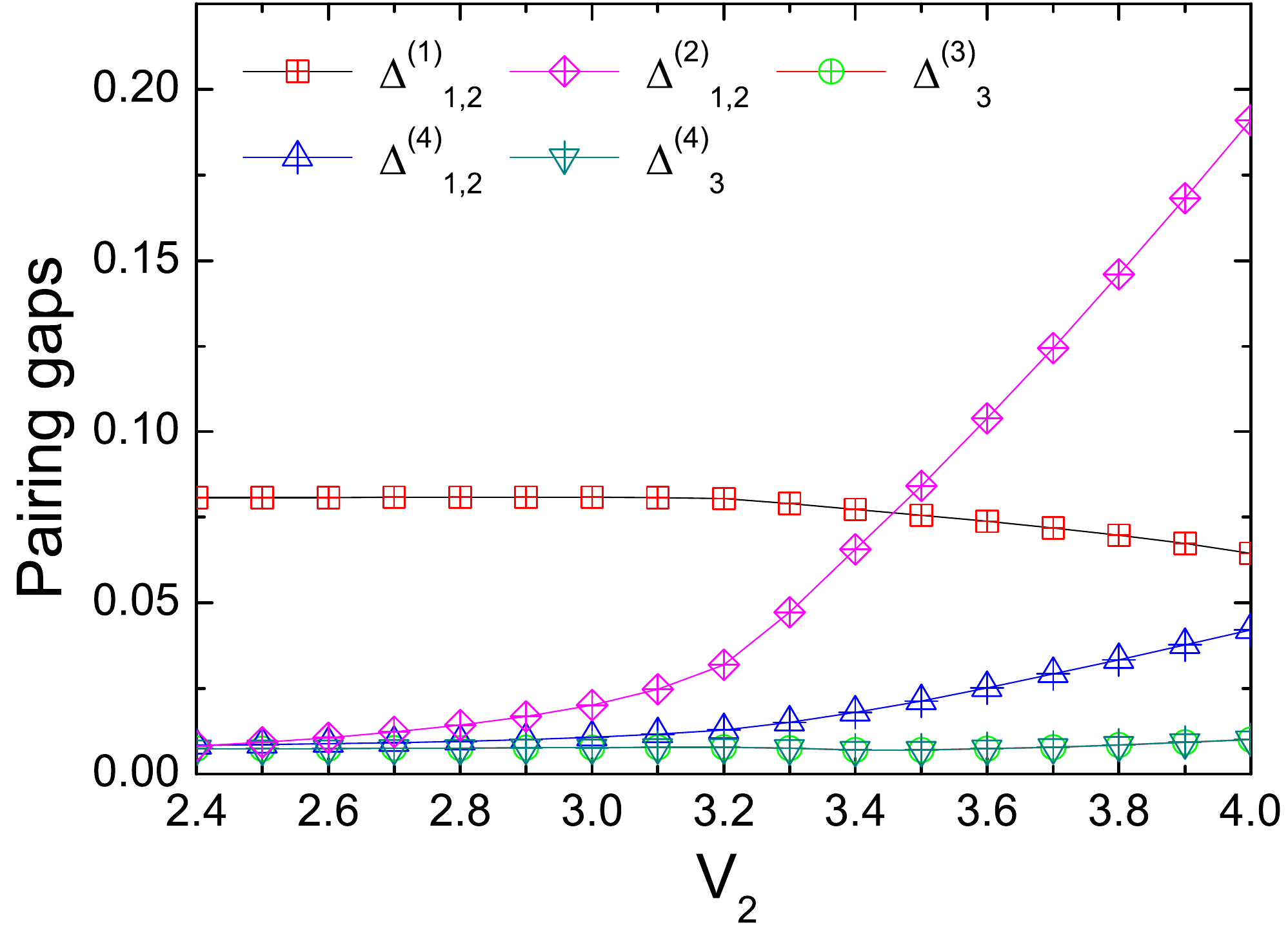}
\caption{ Gaps versus pairing interaction $V_2$ with
$V_1=V_3=V_4=3$. } \label{fig3.4}
\end{figure}

Second, we discuss two important, general results obtained from our model, which are independent of the detailed pairing interaction parameters $V_i$ in $H_I$.
One of these is that the inter-layer pairing always competes with the intra-layer pairing. To demonstrate this more clearly, we switch off $V_3$ and $V_4$
and perform a self-consistent solution with $V_1$ and $V_2$.  The
SC pairing gaps as a function of the interacting
parameters are shown in Fig.\ref{fig3.3}. It is very clear that
the intra-layer pairing reduces while the inter-layer pairing
increases and vice versa. If we turn on all of the inter-layer
pairing interactions, the results are rather similar: while the
different inter-layer pairings can increase simultaneously, the
intra-layer pairing gaps always decrease as the inter-layer ones
increase.  A typical result is presented in Fig.\ref{fig3.4}.
This result qualitatively suggests that a more
quasi two-dimensional SC state is likely better for achieving a higher
$T_c$ since the intra-layer pairing would dominate. So far the highest
$T_c$ in the iron-based superconductors is achieved in the
1111-family. The highest $T_c$ in the 122 family is about 8
degrees lower than the one in the 1111-family \cite{johnston}. Comparing to the 122
family, the 1111 family is much more two-dimensional with much
less dispersion along the c-axis. 

\begin{figure}[bhp]
\includegraphics [angle=0,width=0.8\columnwidth,clip]{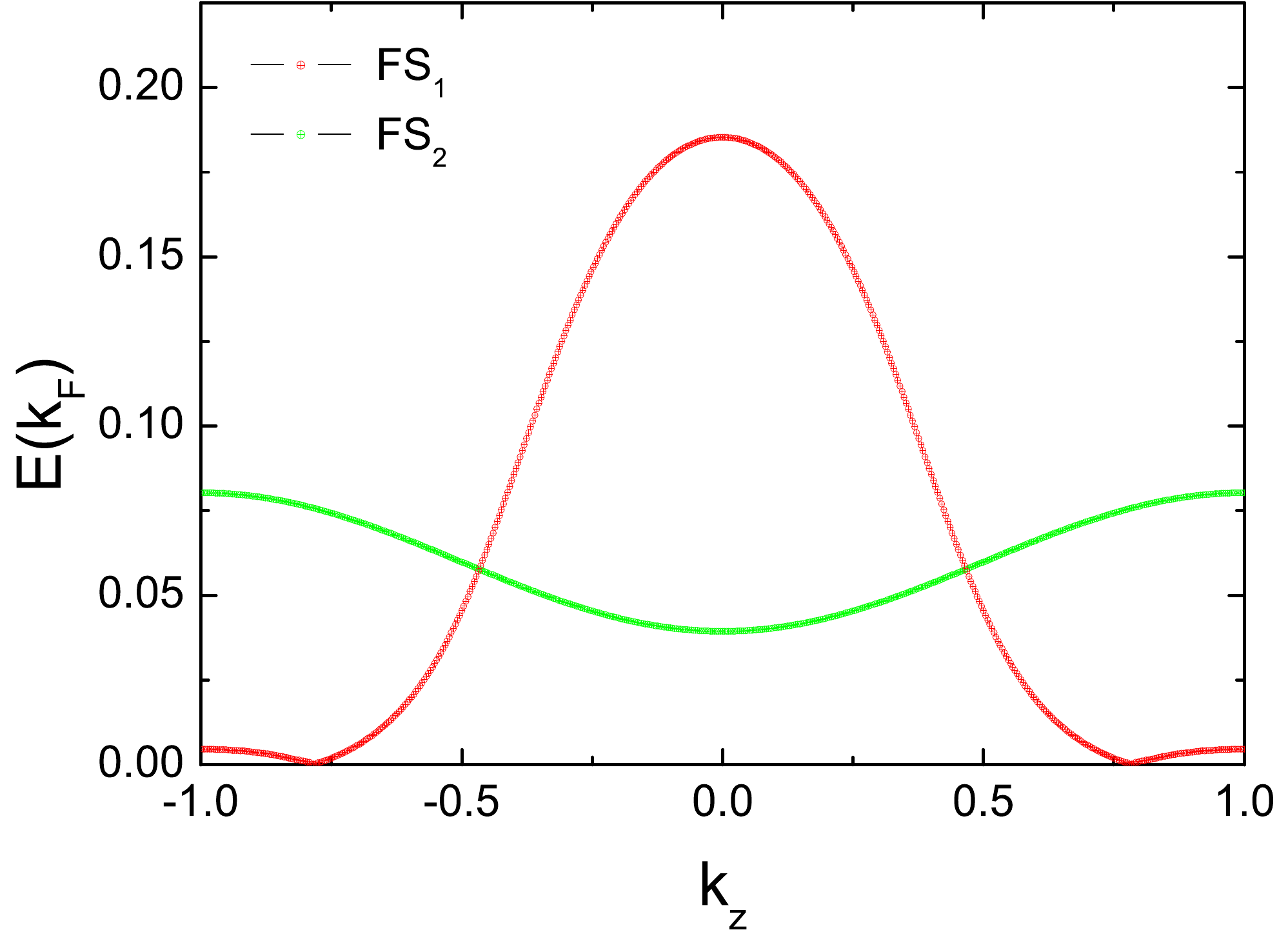}
\caption{ The energy dispersion of the Bogoliubov quasiparticles
at the Fermi surfaces shown in Fig. \ref{fig3.1}. Here $V_1=3,
V_2=3.35, V_3=V_4=3$ with
$\Delta^{(1)}_{1,2}=0.078,\Delta^{(2)}_{1,2}=0.057 ,
\Delta^{(3)}_{3}=0.007,
\Delta^{(4)}_{1,2}=0.016,\Delta^{(4)}_{3}=0.007$.} \label{fig3.5}
\end{figure}

In a multi-orbital model, the relation between the SC pairing parameters
and the energy gap in the low energy single particle excitations can be complicated.
In order to show that the SC state truly develops nodes, we
have to calculate the energy dispersion of the Bogoliubov
quasiparticles at the Fermi surfaces. In Fig.\ref{fig3.5}, we plot
the dispersion of Bogoliubov quasiparticles along the c-axis
with the parameters given by $V_1=3, V_2=3.35, V_3=V_4=3$. There
are two important results. One is that the true nodes can easily
develop on the hole pocket. The other is that
the gap values of the quasiparticles on the hole pocket and
electron pocket are anti-correlated along the c-axis: the gap value on
the hole pocket is larger at $k_z=0$ than at $k_z=\pi$  while on 
the electron pocket, it is smaller at $k_z=0$ than at $k_z=\pi$. This
anti-correlation is a combined result of the inter-layer
pairing and the $S^\pm$-wave symmetry for the intra-layer SC pairing order
parameter which is proportional to $cosk_xcosk_y$ and changes sign between the hole and 
the electron pocket.  This result holds for most of the parameter regions we have
investigated.  Therefore, this inter-layer pairing induced
anti-correlation suggests that ARPES can provide a direct test of
the possible $S^\pm$-wave pairing gap functions in the iron-pnicitide
superconductors. Of course, to detect it, a high energy-resolution
in the ARPES experiments has to be achieved since the c-axis dispersion and the gap
modulation on the electron pockets are not large.

Finally, we note that several previous thermal conductivity
measurements have suggested that the nodes should be on the
electron pockets \cite{linenode2}. A key argument given in \cite{linenode2} is that 
the quaisparticles on the hole
pockets have much lower velocity and heavier mass 
than those on the electron pockets so that the in-plane Fermi velocity
$V_F$ of the hole pockets is too small to explain the observed
residual thermal conductivity. However, this statement is only
partially true. There are three hole pockets centered around the folded Brillouin
zone center. The Fermi velocity on one of the hole pockets is in fact comparable
to that on the electron pockets. ARPES results \cite{hdingtwo} show
that the former is even slightly larger than the later.  This hole
pocket, whose orbital character is even with respect to the
$\Gamma-M$ mirror plane, is exactly the pocket that carries the large
c-axis dispersion. Therefore, the previous thermal conductivity
measurements are consistent with our results for the existence of
gap nodes on the hole pocket.

{\it Conclusion}  We have constructed a model to show how
nodes in the single-particle excitations can emerge in iron-based superconductors.  
The development of the nodes are due to the combined effects of the
increase in the hole pocket size which reduces the SC gap from
intra-layer pairing and the presence of the inter-layer SC
pairing.  This study consistently explains the experimental
observations of the c-axis gap variation in optimally hole-doped
$Ba_{1-x}K_xFe_2As_2$ \cite{ZhangBK,hding} and the nodal behaviors
in $BaFe_2As_{2-x}P_x$ \cite{dlfeng2011}.  We also demonstrated that the
inter-layer and intra-layer pairing generally competes with each
other, and suggested a direct experimental test of the
$S^{\pm}$-wave pairing symmetry through the anti-correlation of the gap modulations
on the hole and electron pockets that can be measured by ARPES. We
believe that our results can also explain the observed nodal behaviors in other
materials such as $KFe_2As_2$ and $AFe_{2-x}Ru_xAs_2$.  A
concrete test of our model will be whether the gap functions
observed in these materials obey Eq. (\ref{eqn3.1}) as the leading contribution
to the quasi-3D SC pairing gap function.

{\it Acknowledgement:}  We thank H. Ding, D. L. Feng, P. C. Dai,
N. L. Wang, H. H. Wen and C. Fang for useful discussions.  Y.H. is
supported by the NSFC (No. 10974167). ZW is supported by DOE DE-FG02-99ER45747.

\end{document}